# Second harmonic generation by radially polarized laser beam propagating in homogeneous plasma

Shivani Aggarwal[1], Dinkar Mishra[1], Saumya Singh[1], Bhupesh Kumar[1*], Pallavi Jha[1,2]

[1]*Department of Physics, University of Lucknow, Lucknow (U.P.) – 226007, India*

[2]*Retired Professor*

**Corresponding author: bhupeshk05@gmail.com*

## Abstract

This study presents an investigation of second harmonic generation (SHG) resulting from interaction of radially polarized laser beam propagating in homogeneous, unmagnetized plasma. Lorentz force and continuity equations have been used to derive the radial and axial current density components. Further, using these densities in the wave equation leads to analytical expressions for the SHG field amplitudes. These amplitudes exhibit periodic oscillations along the propagation axis, characterized by detuning length dependent on plasma density and laser parameters. Radial and axial contributions to SHG are found to be highly enhanced near the beam axis due to the Gaussian beam profile of the laser. The analytical findings are validated using Fourier Bessel Particle-In-Cell (FBPIC) simulations. Notably, unlike linearly or circularly polarized beams which require either inhomogeneous or magnetized plasma, radially polarized beams facilitate efficient harmonic generation in homogeneous unmagnetized plasmas.

***Keywords*:** radially polarized beam, perturbation technique, laser-plasma interaction, second harmonic radiation generation.

## 1.Introduction

When intense laser pulses propagate through plasma, the electrons quiver at velocities approaching the speed of light, leading to strongly nonlinear motion. The resulting laser-plasma interaction is central to many fields, including laser-plasma acceleration (LWFA) [1- 2], attosecond laser pulse generation [3-5], self-focusing [6-7], inertial-confinement fusion [8-10], terahertz radiation generation [11-14], and harmonic generation [15-18]. Sprangle et al. performed the first theoretical evaluation of high-harmonic radiation emitted by a laser beam propagating in homogeneous plasma [19]. The seminal work of Esarey et al. further supported the study of nonlinear laser- plasma interactions for coherent high-frequency generation [20].

Additional advances include higher even and odd-harmonic efficiencies using two- color beams in gases [21], and third-harmonic generation in laser-induced plasma channels in air [22]. Subsequent studies showed that single- and two-color p-polarized beams enhance second and third harmonic generation in homogeneous plasma [23-24]. Brandi et al. analytically demonstrated second-harmonic generation (SHG) from a Hermite-Gaussian laser beam in inhomogeneous plasma [25]. Jha et al. reported SHG driven by a linearly polarized beam in magnetized homogeneous plasma [26], while another analysis found that a circularly polarized beam propagating in obliquely magnetized plasma also yields strong SHG [27]. These studies highlight the higher harmonic generation in underdense plasmas [28].

In recent years, radially polarized laser beams have attracted growing interest as drivers of higher-order harmonic radiation. Theoretical and simulation studies by Wang et al. show that a radially polarized beam, focused by an elliptical mirror, generates efficient SHG [29]. Yew et al. demonstrated that, under tight focusing, both transverse and axial field components of linearly and radially polarized beams contribute to SHG [30]. Theoretical and experimental work has confirmed that radial polarization, together with high numerical aperture focusing, enhances the axial polarization of the generated second harmonic radiation [31-32]. Experimentally, a radially polarized beam with a strong longitudinal field at focus produces significantly stronger SHG than a circularly polarized beam [33]. Mori et al. attributed generation of second harmonic radiation to electrons oscillating at density gradients created by the laser's ponderomotive force [34]. Hashimoto et al. reported that efficiency of SHG from a self-assembled monolayer on platinum is 3.7 times stronger for a radially polarized beam than for a linearly polarized one [35].

This paper presents an analytical formulation of SHG via propagating radially polarized laser pulses through homogeneous underdense plasma and validates its findings through particle-in-cell (PIC) simulations. Section 2 derives the current density and dispersion relations for both the fundamental and second-harmonic frequencies using the Lorentz force and continuity equations. Section 3 presents closed-form expressions for the SHG amplitude and conversion efficiency and illustrates their dependence on propagation distance as well as plasma frequency normalized by laser frequency. Section 4 compares the analytical findings with simulation results, highlighting the effects of pump depletion, laser diffraction, and energy transfer to the harmonic. Section 5 summarizes the principal findings and outlines directions for future work.

## 2. Mathematical formulation

Consider a mildly relativistic, radially polarized laser beam propagating along z-direction in homogeneous plasma having ambient plasma density $n_o \left(= n_e^{(0)}\right)$ . The configuration of electric and magnetic fields are respectively given by,

$$\vec{E} = \sum_{\ell=1}^{2} [\hat{r} E_{o\ell r} \cos(k_\ell z - \ell\omega_o t) \; + \; \hat{z} E_{o\ell z} \sin(k_\ell z - \ell\omega_o t)], \tag{1}$$

and

$$\vec{B} = \sum_{\ell=1}^{2} \hat{\theta} E_{o\ell r} \cos(k_\ell z - \ell\omega_o t), \tag{2}$$

where $E_{o\ell r} \left[= E_o \left(\frac{r}{2r_o}\right) \, exp\left(\frac{-r^2}{r_o^2}\right)\right]$ and $E_{o\ell z} \left[= E_o \, \left(1 - \frac{r^2}{r_o^2}\right) exp\left(\frac{-r^2}{r_o^2}\right)\right]$ are respectively the amplitude of radial and axial components of the laser electric field while $\ell$, $r_{o,}$ $k_\ell$, $\omega_o$ are frequency multiplication factor, beam waist, wave number and fundamental frequency of the laser beam. The plasma electron velocity and density are perturbed by the passage of the radiation field and may be obtained with the help of Lorentz force and continuity equations given by,

$$\frac{\partial \vec{v}}{\partial t} = -\frac{e}{m}\left[\vec{E} + \frac{1}{c}\left(\vec{v} \times \vec{B}\right)\right] - \left(\vec{v}.\vec{\nabla}\right)\vec{v}\,, \tag{3}$$

and

$$\frac{\partial n_e}{\partial t} + \vec{\nabla}.\left(n_e \vec{v}\right) = 0, \tag{4}$$

where $e$, $m$, $n_e$ and $\vec{v}$ are the charge and rest mass, perturbed plasma electron density and velocity respectively.

The propagation of electric component of the laser through plasma is governed by the wave equation

$$\left(\nabla^2 - \frac{1}{c^2}\frac{\partial^2}{\partial t^2}\right)\vec{E} = \frac{4\pi}{c^2}\frac{\partial \vec{J}}{\partial t}\,, \tag{5}$$

where $\vec{J}\,(= -n_e e\vec{v})$ is the plasma electron current density which may be obtained using perturbative technique. The laser and plasma parameters are expanded upto the second order of the radiation field. With the help of Eqs. (1), (2) and (3), first order plasma electron velocities along $\hat{r}$ and $\hat{z}$ directions, are obtained as

$$v_r^{(1)} = \sum_{\ell=1}^{2} a_{o\ell r}\, c \sin(k_\ell z - \ell\omega_o t), \tag{6}$$

$$v_z^{(1)} = -\sum_{\ell=1}^{2} a_{o\ell z}\, c \cos(k_\ell z - \ell\omega_o t), \tag{7}$$

where $a_{o\ell r} = a_{oo}\left(\frac{r}{2r_o}\right)\exp\left(-r^2/r_o^2\right)$, $a_{o\ell z} = a_{oo}\left(1 - \frac{r^2}{r_o^2}\right)\exp\left(-r^2/r_o^2\right)$, and $a_{oo}[= \sum_{\ell=1}^{2}(eE_o/mc\ell\omega_o)] \ll 1$. The equation governing the first order density perturbation is given by

$$\frac{\partial n_e^{(1)}}{\partial t} + \vec{\nabla}.\left(n_e^{(0)}\vec{v}^{(1)}\right) = 0. \tag{8}$$

Using Eqs. (6) and (7), Eq. (8) is solved to give the first order density perturbation as,

$$n_e^{(1)} = - \sum_{\ell=1}^{2} a_{o\ell z} c\, n_o\, k_\ell \frac{1}{(\ell\omega_o)} \cos(k_\ell z - \ell\omega_o t). \tag{9}$$

Substituting Eqs. (6), (7) and (9) into second order Lorentz force equation (3) leads to the radial and axial components of plasma electron velocities as

$$v_r^{(2)} = -\sum_{\ell=1}^{2} a_{o\ell z}\, a_{o\ell r} c^2 \frac{1}{(4\ell\omega_o)} \left(\frac{\ell\omega_o}{c} + k_\ell\right) \sin 2(k_\ell z - \ell\omega_o t) \tag{10}$$

and

$$v_z^{(2)} = \sum_{\ell=1}^{2} \left[-a_{o\ell r}^2 \left(\frac{\ell\omega_o}{c}\right) + a_{o\ell z}^2 k_\ell\right] c^2 \left(\frac{1}{4\ell\omega_o}\right) \cos 2(k_\ell z - \ell\omega_o t). \tag{11}$$

The current density upto the second order of radiation field is given by $\vec{J} = \vec{J}^{(1)} + \vec{J}^{(2)} = -e\left(n_o \vec{v}^{(1)} + n_e^{(1)} \vec{v}^{(1)} + n_o \vec{v}^{(2)}\right)$. Substituting the first and second order plasma electron velocities from Eqs. (6), (7), (10), (11) and first order density from Eq. (9), the radial and axial components of the first and second order current densities are respectively given by,

$$J_r^{(1)} = -n_o ec\, [a_{o1r} \sin(k_1 z - \omega_o t) + a_{02r} \sin(k_2 z - 2\omega_o t)]\,, \tag{12a}$$

$$J_r^{(2)} = n_o ec \left[\frac{a_{o1r} a_{o1z}\, c}{2\omega_o} \left(\frac{3k_1}{2} + \frac{\omega_o}{2c}\right) \sin 2(k_1 z - \omega_o t)\right], \tag{12b}$$

and

$$J_z^{(1)} = n_o ec\, [-a_{o1z} \cos(k_1 z - \omega_o t) - a_{02z} \cos(k_2 z - 2\omega_o t)], \tag{13a}$$

$$J_z^{(2)} = n_o ec \left[\left\{-\frac{3k_1 c}{4\omega_o} a_{o1z}^2 + \frac{1}{4} a_{o1r}^2\right\} \cos 2(k_1 z - \omega_o t)\right]. \tag{13b}$$

It may be noted from Eqns. (12b) and (13b) that the second order current density is oscillating at the second harmonic frequency. Substituting the current density components from Eqs. (12a) and (13a) into the wave equation (Eq. (5)) and equating terms oscillating at the fundamental frequency gives the lowest order dispersion relation for the laser electric field as,

$$c^2 k_1^2 = \omega_o^2 - \omega_p^2. \tag{14}$$

Similarly, the linear dispersion relation for the second harmonic frequency can be obtained as

$$c^2 k_2^2 = 4\omega_o^2 - \omega_p^2. \tag{15}$$

**3. Second harmonic generation**

The amplitude of higher order second harmonic radiation is obtained by substituting the current density (Eqs. (12b) and (13b)) into the wave equation (Eq. (5)) and equating the second harmonic terms. Further, it is assumed that the second harmonic amplitude varies slowly with propagation distance $\left(\frac{\partial^2 a_2}{\partial z^2} \ll k_2 \frac{\partial a_2}{\partial z}\right)$ and that $a_{o1r}$ and $a_{o1z}$ deplete very slightly so that the quantities$a_{o1r}$ , $a_{o1z}$ can be considered to be independent of z. Hence, the evolution of amplitude of $r$ and $z$ components of the second harmonic electric field is given by

$$\frac{\partial(a_{2r})}{\partial z} = i \frac{\omega_p^2}{2k_2\, c} a_{o1r} a_{01z} \left(\frac{3k_1}{2\omega_o} + \frac{1}{2c}\right) \exp\,(i\, \Delta k_2\, z), \tag{16}$$

and

$$\frac{\partial(a_{2z})}{\partial z} = \frac{\omega_p^2}{4\, i\, k_2\, c} \left(-\frac{3k_1}{\omega_o} a_{01z}^2 + \frac{1}{c} a_{01r}^2\right) \exp\,(i\, \Delta k_2\, z), \tag{17}$$

where $\Delta k_2 = 2k_1 - k_2$ is the phase mismatch. While deriving Eqs. (16) and (17), the dispersion relation for the second harmonic frequency has been used. Integrating Eqs. (16) and (17) gives the r and z component of generated second harmonic amplitude as,

$$a_{2r} = i \frac{\omega_p^2}{k_2\, c\, \omega_o\, \Delta k_2} a_{o1r} a_{01z} \left(\frac{3k_1}{2} + \frac{\omega_o}{2c}\right) \exp\,(i\, \Delta k_2\, z/2)\; \sin(\,\Delta k_2\, z/2), \tag{18}$$

and,

$$a_{2z} = i \frac{\omega_p^2}{2\; k_2\, c\, \Delta k_2} \left(\frac{3k_1}{\omega_o} a_{01z}^2 - \frac{1}{c} a_{01r}^2\right) \exp\,(i\, \Delta k_2\, z/2)\; \sin(\,\Delta k_2\, z/2), \tag{19}$$

It may be noted that both $a_{2r}$ and $a_{2z}$ are functions of $r$ as well as $z$. The resultant amplitude of second harmonic radiation generated due to interaction of radially polarized laser beam with unmagnetized plasma is given by,

$$|a_2| = \sqrt{|a_{2r}|^2 + |a_{2z}|^2} = \frac{\omega_p^2}{k_2\ c\ \Delta k_2}\left[\left\{(a_{o1r}a_{o1z})^2\left(\frac{3k_1}{2\omega_o} + \frac{1}{2c}\right)^2\right\} + \left(\frac{3k_1}{2\omega_o}a_{o1z}^2 - \frac{1}{2c}a_{o1r}^2\right)^2\right]^{1/2} \sin(\ \Delta k_2\ z/2). \quad (20)$$

The second harmonic conversion efficiency, defined as the ratio of the intensity of the second harmonic and fundamental frequencies [25], is given by

$$\alpha_2 = \frac{\mu_2|(a_2)^2|}{\mu_1|(a_{o1})^2|}. \quad (21)$$

Here, refractive indices corresponding to the fundamental and second harmonic are $\mu_1 = (ck_1/\omega_o)$ and $\mu_2 = (ck_2/2\omega_o)$ respectively. Substituting the Eq. (20) into Eq. (21), gives second harmonic conversion efficiency as,

$$\alpha_2 = \left(\frac{\omega_p^2}{k_2\ c\ \Delta k_2}\right)^2 \frac{\left[\left\{(a_{o1r}a_{o1z})^2\left(\frac{3k_1}{2\omega_o}+\frac{1}{2c}\right)^2\right\}+\left(\frac{3k_1}{2\omega_o}a_{o1z}^2-\frac{1}{2c}a_{o1r}^2\right)^2\right]}{[(a_{o1r}^2+a_{o1z}^2)]}\sin^2(\ \Delta k_2\ z/2). \quad (22)$$

It may be noted that since $a_{o1r}$ and $a_{o1z}$ are function of r, the conversion efficiency for a given set of laser and plasma parameters varies with $r$ as well as $z$.

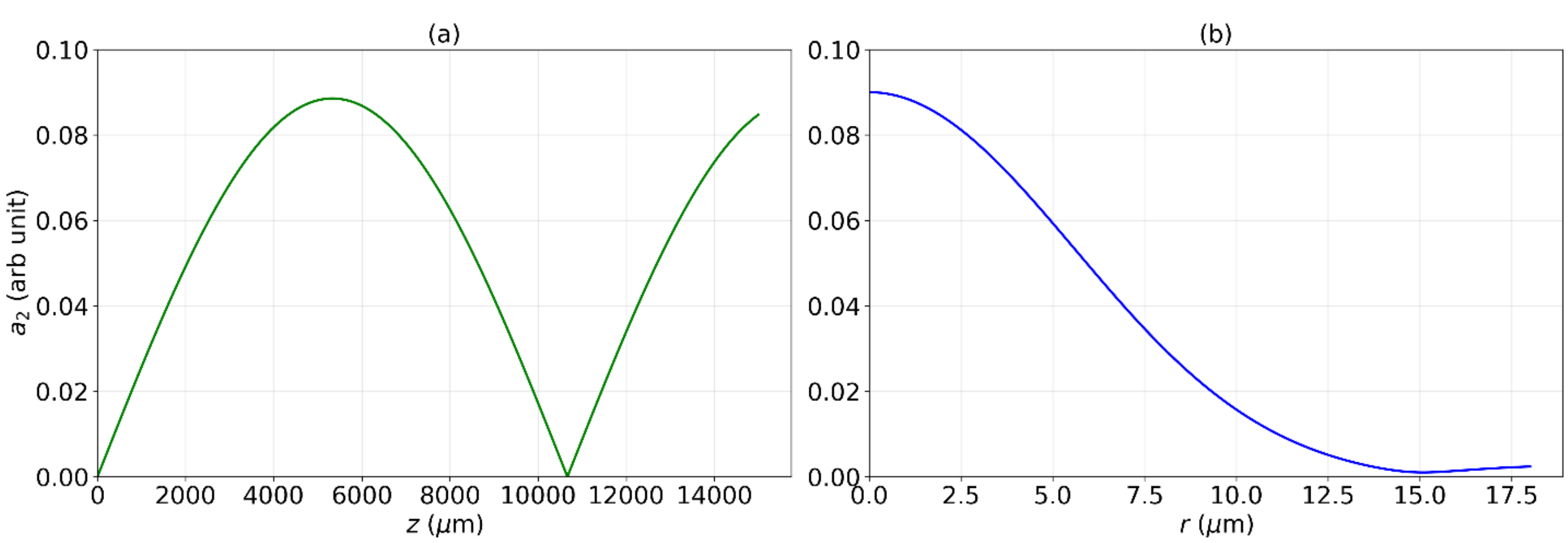

Figure 1. Variation of second harmonic amplitude $a_2$ (a) with respect to propagation distance z and (b) with respect to radial distance r for $a_{oo} = 0.3$, $\lambda_o = 0.8\,\mu m, r_o = 15\,\mu m$ , $\omega_p/\omega_o = 0.01$, and $\omega_p = 0.23 \times 10^{14} sec^{-1}$ ($n_o$ = 1.75×$10^{24}$ $m^{-3}$).

Analytical results are graphically depicted for the laser and plasma parameters $a_{oo} = 0.3$, $\lambda_o = 0.8\,\mu m, r_o = 15\,\mu m$ , $\omega_p/\omega_o = 0.01$ ,and $\omega_p = 0.23 \times 10^{14} sec^{-1}$ ($n_o$ = 1.75×$10^{24}$ $m^{-3}$). Figure 1(a) presents the analytical variation of the generated second harmonic amplitude $a_2$, derived from Eq. (20), with z at $r = 1\,\mu m$. It is seen that the amplitude modulates periodically with z, reaching a peak amplitude of approximately 0.09 at z ≈ 5,333 μm before decreasing due to phase mismatch. The amplitude of second harmonic radiation is maximum for $\Delta k_2 z_d/2 = \pi/2$ where $z_d = \pi/\Delta k_2$ is the detuning length. The oscillatory behavior corresponds to the coherence length, calculated analytically, which governs the periodic energy exchange between the fundamental and second harmonic fields. Similarly, the variation of amplitude with r at $z = 5{,}333\mu m$. Figure 1(b) shows that $|a_2|$decreases monotonically with increasing r, due to the combined effect of Gaussian beam profile and dependence of the laser amplitude with significant SHG confined near the axis. Figure 2 presents the variation of analytical second harmonic generation efficiency $\alpha_2$ with respect to axial and radial positions for the same parameters as in Fig. 1. Figure 2 (a) exhibits a smooth sinusoidal modulation of $\alpha_2$ along z, reaching a peak value of ~ 0.1 while Fig. 2(b) shows a radial variation of $\alpha_2$ having maximum value (~ 0.1) near the axis.

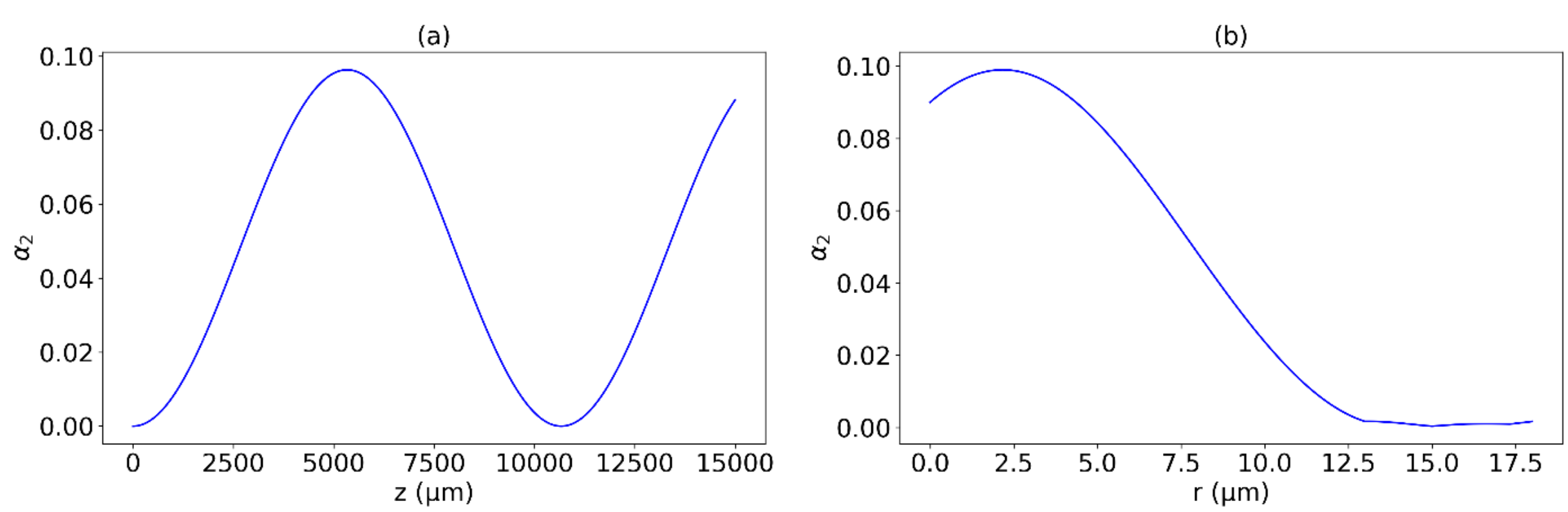

Figure 2. Variation of second harmonic efficiency $\alpha_2$ (a) along z (b) along r for laser and plasma parameters as $a_{oo} = 0.3$, $\lambda_o = 0.8\,\mu m$, $r_o = 15\,\mu m$ , $\omega_p/\omega_o = 0.01$ ,and $\omega_p = 0.23 \times 10^{14} sec^{-1}$ ($n_o = 1.75\times10^{24}$ $m^{-3}$).

## 4.Simulation Results

In order to simulate second harmonic generation in homogeneous and unmagnetized plasma driven by a radially polarized laser, fully electromagnetic particle-in-cell (PIC) simulations were performed using the Fourier Bessel Particle in cell (FBPIC) code in a quasi 3D - cylindrical symmetry. The simulation domain extended from z = −10 μm to z = 120 μm along the longitudinal axis with $N_z$ = 1800 grid cells, and up to r = 20 μm radially with $N_r$ = 150 grid points. Three azimuthal Fourier modes ($N_m$ = 3) were retained to accurately capture the cylindrical symmetry of the radially polarized beam and associated harmonics. A moving window propagating at the speed of light was employed to efficiently follow the laser beam through the plasma. The temporal resolution was determined by the Courant condition and set as $\Delta t = (z_{max} - z_{min})/(N_z c) = 2 \times 10^{-16}$ sec, ensuring numerical stability.

The pre-ionized plasma density ramp was initialized with a uniform electron density of $n_o$ = $1.75\times10^{24}$ $m^{-3}$ $\left(\omega_p = (4\pi n_e e^2/m)^{1/2}\right)$, extending from z = 120 μm to z = 250 μm, and radially confined within r ≤ 18 μm. To model the particle distribution in the simulation domain, 2 particles per cell were loaded along z and r, and 4 in the azimuthal (θ) direction. The continuous laser beam having laser strength parameter of $a_{oo}$ = 0.3, with a beam waist of $r_o$ = 15 μm, and wavelength $\lambda_o = 0.8\,\mu m$ was focused at z = 100 μm. This configuration enabled high-resolution study of second harmonic field components under realistic experimental conditions.

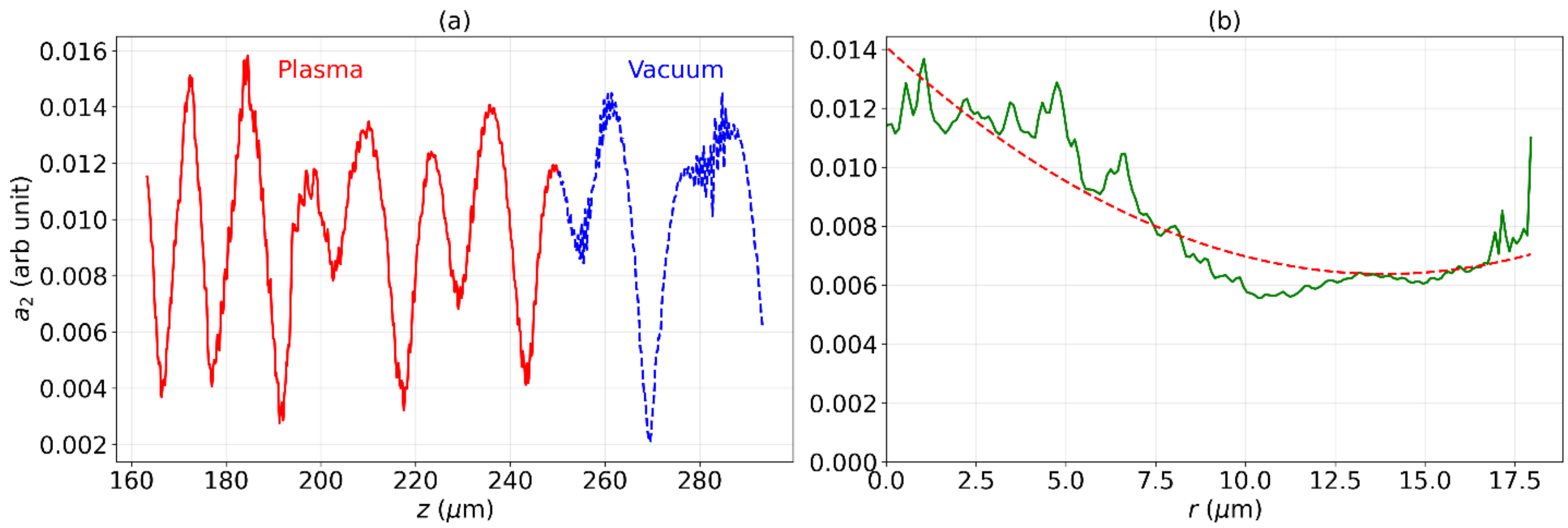

Figure 3. Variation of the second harmonic generation (SHG) amplitude $a_2$ for $a_{oo} = 0.3$, $\lambda_0 = 0.8$ $\mu$m, beam waist $r_0 = 15$ μm, and $\omega_p/\omega_o = 0.01$. (a) $a_2$ versus axial distance $z$ at fixed $r = 1$ $\mu$m, showing oscillatory behavior in the plasma region (solid) and in vacuum (dashed). (b) $a_2$ versus radial position r at fixed $z = 5{,}333$ $\mu$m, where the solid curve shows the computed profile and the dashed curve represents its best fit.

Figure 3 illustrates the variation of the second harmonic generation (SHG) amplitude $a_2$ with respect to the axial position ($z$) and radial position ($r$). In Fig. 3(a), the SHG amplitude is plotted along the z-axis at a fixed radial distance of $r = 1$ μm near the axis. The ratio of plasma frequency to laser frequency $(\omega_p/\omega_o)$ was considered to be 0.01. The solid curve represents the plasma region, while the dashed curve marks the vacuum region beyond $z = 250$ μm. The SHG amplitude exhibits oscillatory behavior in the plasma region as well as in the vacuum, with a peak amplitude of $a_2 \sim 0.014$. This oscillation arises on account phase-mismatch between frequencies of the laser beam and generated second harmonic radiation. In Fig. 3(b), the radial dependence of $a_2$ is shown at a fixed axial position $z = 5{,}333$ μm. The solid curve corresponds to the computed SHG amplitude, and the dashed line represents a second-order polynomial best fit to the simulation data. The SHG amplitude peaks near the axis with a maximum value of $a_2 \sim 0.014$ and decreases monotonically with increasing r, reflecting the radial intensity profile of the SHG radiation. It may be noted that generated second harmonic radiation is also propagating in vacuum $(z > 250\ \mu m)$ and may be detected outside the plasma boundary.

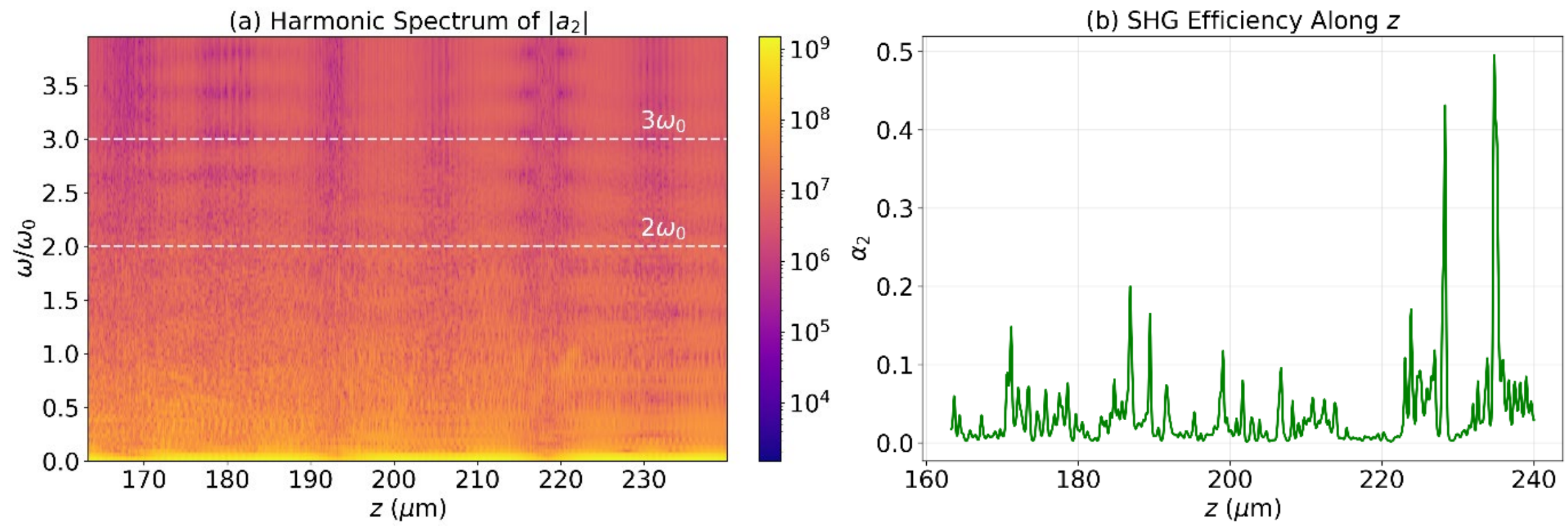

Figure 4. (a) Harmonic spectrum of the amplitude ($a_2$) as a function of axial position z and (b) SHG efficiency ($\alpha_2$) as a function of axial position $z$ and its best fit at $r = 1\ \mu m$.

Figure 4 presents the harmonic spectrum and efficiency of generated second harmonic radiation for a radially polarized laser beam propagating in unmagnetized plasma. Fig. 4(a) shows the amplitudes of the spectrum of generated harmonics and the possibilities of enhancent second and higher harmonic frequencies. Figure 4(b) shows the efficiency ($\alpha_2$) derived from the simulation data. It may be seen that the efficiency of the generated SHG follows the same trend as obtained in analytical depiction (Fig. 2).

Comparing Figs. 1and 3, it may be noted that amplitude of the generated second harmonics derived via analytical study is greater than that obtained via simulation. In addition, detuning length obtained via analytical calculation (simulation study) is ~5333 $\mu m$ (~ $5\mu m$). These discrepancies may be attributed to the realistic evolution of the laser beam neglected in the analytical study while being taken into consideration in the simulation study. As the laser beam propagates in plasma, it undergoes diffraction and energy depletion. These effects reduce the amplitude of the generated harmonics as seen via simulation results. These complex nonlinear dynamic effects contribute to much shorter detuning length than evaluated in analytical formulation.

## 5. Summary and Conclusions

The present study investigates second harmonic generation (SHG) resulting from the interaction of a radially polarized laser beam with homogeneous and unmagnetized plasma. Perturbation technique is used to obtain the current density components with the help of Lorentz force and continuity equations. The corresponding dispersion relation for the second harmonic frequencies is derived. Substitution of the radial and axial components of the current densities into the wave equation yield the respective SHG field amplitudes. The resultant amplitude of the second harmonic generated behind the laser pulse is obtained by combining the radial and axial components. Analytical findings are compared and validated with quasi-3D particle-in-cell (PIC) simulations carried out using the FBPIC code. The maximum second-harmonic amplitude obtained analytically (via simulation) is approximately 0.09 (0.014). The discrepancy in analytical and simulation results arises because the analytical formulation neglects realistic beam evolution effects such as diffraction and energy depletion, that are inherently captured in simulations. As the laser propagates through plasma, these nonlinear dynamical processes reduce the harmonic amplitude and lead to a significantly shorter detuning length than predicted analytically. Furthermore, the generated second harmonic is observed to propagate into vacuum, implying that it can be detected beyond the plasma boundary. The novelty of this study lies in demonstrating that radially polarized laser beams provide an efficient mechanism for generation of radially polarized higher-harmonic frequencies in homogeneous, unmagnetized plasma.

**Data Availability Statement**

Data used in the study is available on reasonable request to the Authors.